\documentclass[aps,pra,reprint, amsmath, amssymb,superscriptaddress,nofootinbib]{revtex4-1}

\usepackage{bm}
\usepackage[retainorgcmds]{IEEEtrantools}
\usepackage{graphicx}
\usepackage{mathrsfs}
\usepackage{amsmath}
\usepackage{amssymb}
\usepackage{color}
\usepackage{amsfonts}
\usepackage{times,txfonts}
\usepackage{nicefrac}
\usepackage[colorlinks=true,linkcolor=blue,urlcolor=blue,citecolor=blue,pdfusetitle]{hyperref}

\newcommand{\nn}{\bar{n} + {1}/{2}}
\newcommand{\tr}{\text{tr}}

\begin{document}

\title{Non-linear Onsager relations for Gaussian quantum maps}
\date{\today}
\author{Domingos S. P. Salazar}
\affiliation{Unidade de Educa\c c\~ao a Dist\^ancia e Tecnologia,
Universidade Federal Rural de Pernambuco,
52171-900 Recife, Pernambuco, Brazil}
\author{Gabriel T. Landi}
\email{gtlandi@if.usp.br}
\affiliation{Instituto de F\'isica da Universidade de S\~ao Paulo,  05314-970 S\~ao Paulo, Brazil.}

\begin{abstract}

Onsager's relations allow one to express the second law of thermodynamics in terms of the underlying associated currents. 
These relations, however, are usually valid only close to equilibrium. 
Using a quantum phase space formulation of the second law, we show that open bosonic Gaussian systems also obey a set of Onsager relations, valid arbitrarily far from equilibrium.  
These relations, however, are found to be given by a more complex non-linear function, which reduces to the usual quadratic form close to equilibrium. 
This non-linearity implies that far from equilibrium, there exists a fundamental asymmetry between entropy flow from system to bath and vice-versa. 
The ramifications of this for applications in driven-dissipative quantum optical setups are also discussed. 

\end{abstract}
\maketitle{}

\section{Introduction}

In the context of irreversible thermodynamics, when a system is pushed away from equilibrium it responds by developing currents (of heat, particles, etc.), which cause it to eventually re-equilibrate. 
The entire process can therefore be understood as an interplay between thermodynamic forces (affinities) $f_i$, such as temperature gradients, and the corresponding system response, in the form of currents $\phi_i$ from the system to the environment. 
The entropy production rate,  dictating how far the system is from equilibrium, is \cite{Callen1985}
\begin{equation}\label{IT_Pi_gen}
    \Pi = \sum\limits_i f_i \phi_i.
\end{equation}
For instance, in a system supporting currents of energy and particles, one would have $\phi_1 = \dot{U}$ and $\phi_2 = \dot{N}$, the rate of change of the internal energy and particle number. 
The corresponding affinities will then be the temperature gradient, $f_1 = \delta(1/T)$, and the chemical potential gradient $f_2 = \delta(\mu/T)$.
When the affinities are small, the response tends to be linear, 
\begin{equation}\label{linear_response}
    \phi_i = \sum_i \textrm{L}_{ij} f_j,
\end{equation} 
where the coefficients $\textrm{L}_{ij}$ form what is known as the Onsager matrix \cite{Onsager1931}.
According to Onsager's reciprocity theorem, $L$ is symmetric and positive semidefinite. 
Within this linear response regime, 
Eq.~(\ref{IT_Pi_gen}) becomes a quadratic form 
\begin{equation}
\label{Onsagers}
    \Pi = \sum\limits_{ij} \textrm{L}_{ij} \phi_i \phi_j.
\end{equation}
Whence, in irreversible thermodynamics, linear response is characterized by a quadratic relation between currents and entropy production \cite{DeGroot1961}.
Far from equilibrium, this quadratic form no longer holds and no general connection exists between entropy production and flux.

The  results above apply to macroscopic thermodynamic systems. 
But they can also be extended to  the microscopic realm, both classical (stochastic thermodynamics) and quantum.
Classical systems are usually modeled using either Fokker-Planck or Pauli master equations \cite{VanKampen2007,Seifert2005,Tome2010,VandenBroeck2010,Seifert2012}. 
This allows one to identify generalized affinities and currents at the stochastic level~\cite{Esposito2010d}, such that the entropy production can still be decomposed in the form Eq.~(\ref{IT_Pi_gen}) and Onsager relations~(\ref{Onsagers}) continue to hold close to equilibrium. 
Far from equilibrium, on the other hand, stochastic systems are found to obey Fluctuation Theorems~\cite{Evans1993,*Evans1994,Gallavotti1995b,*Gallavotti1995,Jarzynski1997,*Jarzynski1997a,Crooks1998,*Crooks2000}.
These are more general and imply Onsager's relation close to equilibrium~\cite{Andrieux2004}.

Conversely, irreversible thermodynamics in the quantum regime is usually studied using either quantum master equations~\cite{Spohn1978,Breuer2003,Breuer2007} or non-equilibrium Green's functions~\cite{Stefanucci,Yamamoto2015}. 
The latter is perhaps the case where Onsager's relations find most applications, specially in the field of thermoelectrics~\cite{Bell2008}. 
The Onsager cross coefficients $L_{12}$ and $L_{21}$ are related to the Seebeck and Peltier effects, which are the basis for several technological applications. 
They are also related to the output power when a thermoelectric is interpreted as an autonomous quantum heat engine~\cite{Benenti2017a,Josefsson2018,Chiaracane2019}. 

Onsager's relations in quantum master equations, on the other hand, have been much less explored~\cite{Lendi2001,Guimar2016}.
Some simple scenarios, such as Davies maps~\cite{Heidelberg1987,Breuer2007}, can actually be converted into Pauli master equations, so that the stochastic thermodynamics formalism applies. 
More general cases, however, can quickly run into serious difficulties, particularly due to quantum coherent effects. 
For instance, in Ref.~\cite{Miller2019} the authors have shown that even in the case where there is only one associated flux (so that Onsager's matrix would be $1\times 1$), the entropy production will contain a non-trivial contribution due to quantum coherence.~\footnote{The results of~\cite{Miller2019} are phrased in terms of work quantities, but can be rephrased in terms of entropy production}

In this paper we analyze  Onsager's relations in the context of continuous-variable bosonic systems, where thermodynamics can be constructed solely in terms of quantum phase space~\cite{Adesso2012,Deffner2013a,Pigeon2015,Santos2017b,Brunelli2016a,Friis2017,Macchiavello2020}.
Our main result is to show that these systems also obey an Onsager relation. However, unlike Eq.~(\ref{Onsagers}), the entropy production and flux are related by a more complicated non-linear function. 
To  elucidate this, we anticipate  the result for the  simplest possible scenario of a single  mode relaxing to equilibrium. In this case our Onsager relation reads
\begin{equation}\label{simplest}
\Pi = \frac{\phi^2}{\phi + \gamma},
\end{equation}
where $\gamma>0$ is a constant related to the underlying  dynamics.
This result is valid arbitrarily far from equilibrium and reduces to Eq.~(\ref{Onsagers}) in the linear response regime ($\phi \ll \gamma$).
The non-linear structure of Eq.~(\ref{simplest}), however, implies that $\Pi$ is not an even function of the flux; i.e.,
\begin{equation}
    \Pi[\phi] \neq \Pi[-\phi]. 
\end{equation}
An even dependence of $\Pi$ on $\phi$ is a hallmark of classical Onsager relations. 
It implies that the entropy production does not depend on the direction of the flow, only on its magnitude. 
Eq.~(\ref{simplest}), however, shows that far from equilibrium this asymmetry is fundamentally broken. In fact, one finds that 
\begin{equation}\label{asymmetry}
    \frac{1}{\Pi[\phi]} - \frac{1}{\Pi[-\phi]} = \frac{2}{\phi},
\end{equation}
a result which, shown below, is actually general. 

\section{The model}

We consider here a system of $L$ bosonic modes characterized by operators $R= (a_1, \ldots, a_L)$ satisfying the usual algebra $[a_i, a_j^\dagger] = \delta_{ij}$. 
We assume that the first moments are zero and define the covariance matrix (CM) $\Theta_{ij} = \frac{1}{2} \langle \{ R_i, R_j^\dagger\} \rangle$. 
Classical CMs are only restrcited to be positive definite; the Heisenberg uncertainty principle, however, imposes the stronger (bona fide) constraint~\cite{Serafini2017}
\begin{equation}\label{bona_fide}
    \Theta - \frac{i}{2} \Omega \geq 0, 
\end{equation}
where $\Omega = (-i\sigma_z)^{\oplus L}$ is the symplectic form for our CM (with $\sigma_z$ being the usual Pauli matrices). 

Our focus will be on Gaussian states and Gaussian-preserving maps \cite{Serafini2017}.
This encompasses a multitude of experimentally relevant situations, such as optomechanics \cite{Paternostro2006,Paternostro2007,Aspelmeyer2014}, ultra-cold atoms~\cite{Baumann2010}, non-linear optics~\cite{Barbosa2018} and others. 
We also assume a continuous time Markovian evolution, which includes both Lindblad as well as quantum Langevin dynamics. 
The CM in this case evolves according to the Lyapunov equation
\begin{equation}\label{lyap}
    \frac{d \Theta}{d t} = W \Theta + \Theta W^\dagger + F.
\end{equation}
Here $W =  \Omega H - \Gamma/2$ is a matrix composed of a Hamiltonian part $\Omega H$ and a dissipative part $\Gamma$.
We shall assume, for the sake of concreteness, that $\Gamma$ has the form  $\Gamma = \text{diag} (\gamma_1, \gamma_1, \ldots, \gamma_L, \gamma_L)$, for $\gamma_i \geq 0$.

The matrix $F$ in Eq.~(\ref{lyap}) is known as the diffusion matrix. 
In classical stochastic processes, the only restriction imposed on $F$ is positive semi-definiteness~\cite{Hespanha}. 
For quantum process, however, one must ensure that the map is completely positive and trace preserving (CPTP).
A general Gaussian map of the form  $\Theta \to X \Theta X^\dagger + Y$
is CPTP provided the matrices $X$ and $Y$ satisfy~\cite{Lindblad2000}
\begin{equation}\label{CPTP}
    \frac{i}{2} (X\Omega X^\dagger - \Omega) + Y \geq 0.
\end{equation}
Integrating Eq.~(\ref{lyap}) over an infinitesimal interval yields a Gaussian map with $X = 1 + W d t$ and $Y = F dt$. Eq.~(\ref{CPTP}) therefore implies that $\frac{i}{2} (W\Omega + \Omega W^\dagger) + F \geq 0$. 
It is convenient to parametrize
\begin{equation}\label{F}
    F = \frac{1}{2}(\Gamma Q + Q \Gamma). 
\end{equation}
Eq.~(\ref{CPTP}) then implies the constraint
\begin{equation}
    Q - \frac{i}{2} \Omega \geq 0. 
\end{equation}
In words, the Lyapunov equation~(\ref{lyap}) will produce a genuine quantum Gaussian evolution if $Q$ represents a valid Gaussian CM [c.f.~Eq.~(\ref{bona_fide})]. 
If the Hamiltonian part of $W$ is zero, then the steady-state of Eq.~(\ref{lyap}) will be precisely $\Theta(t\to\infty) = Q$. 
When there are Hamiltonian terms, however, the steady-state will in general differ from $Q$ and will often be a non-equilibrium state. 
For simplicity, we will henceforth assume that $[Q,\Gamma]=0$, as this is almost always the case. Eq.~(\ref{F}) then simplifies to $F = \Gamma Q$.

\section{Entropy production rate}

Gaussian systems are naturally characterized by the R\'enyi-2/Wigner entropy, which is given by~\cite{Adesso2012} 
\begin{equation}\label{entropy}
    S(\Theta) = \frac{1}{2} \ln |2\Theta|.
\end{equation}
Using the relation $\frac{d}{dt} \ln |\Theta| = \tr\big(\Theta^{-1} \frac{d\Theta}{dt}\big)$, together with the Lyapunov Eq.~(\ref{lyap}), one  finds that 
\begin{equation}\label{dSdt}
    \frac{dS}{dt} 
     = \frac{1}{2} \tr \bigg( \Gamma Q \Theta^{-1} -  \Gamma \bigg) .
\end{equation}
Due to the interaction with the bath, the entropy of the system may either decrease or increase, so that $dS/dt$ does not have a definite sign.
The part of the change in entropy which is always non-negative is the entropy production rate, which is given by 
\begin{equation}\label{Pi0}
    \Pi = \frac{dS}{dt} + \Phi \geq 0, 
\end{equation}
where $\Phi$ is called the entropy flux rate, from the system to the environment. 
As shown in~\cite{Landi2013b,Santos2017b,Malouf2018a}, for general Lyapunov equations $\Phi$ can be written as 
\begin{equation}\label{flux}
    \Phi = \frac{1}{2} \tr\bigg( \Gamma \Theta Q^{-1}  - \Gamma \bigg\}. 
\end{equation}
Combining this with Eq.~(\ref{dSdt}) then yields the entropy production rate
\begin{equation}\label{Pi}
    \Pi = \frac{1}{2} \tr\bigg( \Gamma (Q - \Theta)(\Theta^{-1} - Q^{-1}) \bigg) \geq 0.
\end{equation}
The non-negativity of $\Pi$ can be made apparent by writing $(Q - \Theta)(\Theta^{-1} - Q^{-1}) = M Q M^\dagger \Theta^{-1}$, where $M = 1 - \Theta Q^{-1}$, which is manifestly positive definite and therefore so is the trace.
The entropy production~(\ref{Pi}) serves a natural quantifier of how far the system is from equilibrium, in the sense that it clearly measures the distance between the CM $\Theta$ and the bath-imposed CM $Q$. 

\section{Non-linear Onsager relations}

We now show how Onsager's relations emerge in our treatment, in the form of a non-linear expression valid arbitrarily far from equilibrium. 
The matrix nature of Eqs.~(\ref{flux}) and (\ref{Pi}) naturally suggest that we define a \emph{flow matrix}
\begin{equation}\label{upsilon}
    \Upsilon = \Gamma^{1/2} \Theta  Q^{-1}\Gamma^{1/2} - \Gamma,
\end{equation}
and a \emph{production matrix}
\begin{equation}\label{xi}
    \Xi = \Gamma  (Q - \Theta)(\Theta^{-1} - Q^{-1}). 
\end{equation}
Eqs.~(\ref{flux}) and (\ref{Pi}) are then written as 
\begin{equation}\label{Pi_Phi_traces}
\Phi = \frac{1}{2} \tr(\Upsilon),
\qquad
\Pi = \frac{1}{2} \tr(\Xi). 
\end{equation}
The two matrices are actually related to each other. 
Using standard matrix algebra, one finds\footnote{When $\Gamma$ is not full rank, $\Gamma^{-1}$ is to be interpreted as the Moore-Penrose generalized inverse~\cite{Landi2013b}.} 
\begin{equation}\label{Xi_Upsilon}
    \Xi = \Upsilon \; \Gamma^{-1} \Upsilon (\Gamma + \Upsilon)^{-1} \Gamma.
\end{equation}
This is the matrix version of the generalized Onsager relation. 
It implies that the entropy production can be written solely as 
\begin{equation}\label{onsager_gen}
    \Pi = \frac{1}{2} \tr\bigg\{  \; \Upsilon \; \Gamma^{-1} \; \Upsilon \;(\Gamma + \Upsilon)^{-1}\Gamma\bigg\}, 
\end{equation}
which is a function \emph{only} of the flow matrix $\Upsilon$ and the damping rate $\Gamma$. 
This is the main result of this paper: a non-linear Onsager relation valid arbitrarily far from equilibrium. 
The linear response regime is recovered when $\Gamma + \Upsilon \simeq \Gamma$, in which case Eq.~(\ref{onsager_gen}) simplifies to 
\begin{equation}
    \Pi = \frac{1}{2} \tr\bigg( \Gamma^{-1} \Upsilon^2\bigg),
\end{equation}
which is the traditional quadratic Onsager relation.

The physics behind Eq.~(\ref{onsager_gen}) can be made more transparent by considering the particular case where $\Gamma = \gamma I$ is proportional to the identity matrix (of dimension $2L$). 
This is true for a single mode or for multiple modes with identical damping rates. 
In this case we get
\begin{equation}
    \Pi = \frac{1}{2} \tr \bigg( \frac{\Upsilon^2}{\gamma I + \Upsilon}\bigg).
\end{equation}
Let us further assume that  the target state $Q$ of the Lyapunov equation~(\ref{lyap}) is a thermal state of the form $Q = (\bar{n}+\nicefrac{1}{2}) I$. 
If all modes initially start in a thermal state with the same occupation, then the entire evolution of the covariance matrix will be trivially given by $\Theta(t) = \theta_t I$, where $\theta_t = \langle a_i^\dagger a_i \rangle + \nicefrac{1}{2}$ (independent of $i$). The dynamics of $\theta_t$ is given by the Lyapunov equation~(\ref{lyap}) and reads
\begin{equation}
    \frac{d\theta_t}{dt} = \gamma ((\bar{n}+\nicefrac{1}{2})-\theta_t), 
\end{equation}
which is in the form of the so-called law of cooling.
The flow matrix in this case simplifies to 
\begin{equation}
    \Upsilon = \gamma \bigg( \frac{\theta_t}{\bar{n}+\nicefrac{1}{2}} - 1\bigg) I  = \frac{\Phi}{L} I. 
\end{equation}
Whence, the entropy production becomes
\begin{equation}\label{simplest2}
        \Pi[\Phi] =  L \frac{(\Phi/L)^2}{\gamma + \Phi/L},
\end{equation}
which is Eq.~(\ref{simplest}) with a flow per mode $\phi = \Phi/L$. 

As already touched upon in the introduction, the fundamental new feature of the non-linear Onsager relation is the asymmetry with respect to positive or negative entropy flows. 
The inverse of the production matrix~(\ref{xi}) has the form
\begin{equation}
    \Xi^{-1} = \Upsilon^{-1} \Gamma \Upsilon^{-1} + \Upsilon^{-1}.
\end{equation}
Whence, one readily finds 
\begin{equation}\label{odd}
    \Xi[\Upsilon]^{-1} - \Xi[-\Upsilon]^{-1} = 2 \Upsilon^{-1},
\end{equation}
which is the matrix version of~(\ref{asymmetry}). 
It shows how the parity of $\Xi[\Upsilon]$ is broken far from equilibrium. 
Note also how this asymmetry depends \emph{only} on the flow matrix $\Upsilon$ and not $\Gamma$. 

\section{Applications}

\subsection{Optical parametric oscillator}

As a first application, we consider an optical parametric oscillator described by the Hamiltonian 
\begin{equation}
    \mathcal{H} = - \frac{i \chi}{2} (a^{\dagger 2} - a^2), 
\end{equation}
and subject to a heat bath at occupation $\bar{n}$. 
The matrices $W$ and $F$ in Eq.~(\ref{lyap}) in this case read 
\begin{equation}
    W = - \begin{pmatrix} \gamma & \chi \\[0.2cm] \chi & \gamma/2 \end{pmatrix}, 
    \qquad 
    F = \gamma(\bar{n}+\nicefrac{1}{2}) I_2.
\end{equation}
We focus on the steady-state, which is a solution of $W\Theta_\text{ss} + \Theta_\text{ss} W^\dagger = - F$. 
It reads
\begin{equation}
    \Theta_\text{ss} = \frac{\nn}{\gamma^2 - 4 \chi^2} \begin{pmatrix} \gamma^2 & -2  \gamma \chi \\[0.2cm]
    -2\gamma \chi & \gamma^2 \end{pmatrix}.
\end{equation}
The steady-state exists provided $4 \chi^2 <\gamma^2$; above this threshold the problem becomes unstable and $\Theta(t)$ diverges. 
The steady-state flow matrix~(\ref{upsilon}) is given by 
\begin{equation}
    \Upsilon_\text{ss} = \frac{2\gamma \chi}{\gamma^2 - 4\chi^2} \begin{pmatrix} 2\chi & - \gamma \\[0.2cm] - \gamma & 2\chi \end{pmatrix}. 
\end{equation}
It vanishes when either $\gamma = 0$ (meaning we uncouple the system from the bath) or when $\chi = 0$ (in which case the system relaxes towards thermal equilibrium. 
Remarkably, note that $\Upsilon$ (and hence the steady-state entropy flux) is completely independent of the bath temperature $\bar{n}$. 
The production matrix~(\ref{xi}) is similarly given by 
\begin{equation}
    \Xi_\text{ss} = \frac{4\gamma\chi}{\gamma^2 - 4\chi^2} \begin{pmatrix} 
    \chi & - 2\chi^2/\gamma \\[0.2cm] - 2\chi^2/\gamma & \chi \end{pmatrix}. 
\end{equation}
Using Eq.~(\ref{Pi_Phi_traces}) one then finds that, in the steady-state, 
\begin{equation}
    \Pi_\text{ss} = \Phi_\text{ss} = \frac{4\gamma \chi^2}{\gamma^2 - 4\chi^2}.
\end{equation}
The entropy production differs from zero due to a competition between the damping rate $\gamma$, which tries to push the system towards the vacuum, and the interaction $\chi$, which tends to squeeze the mode. 

\subsection{Detuned squeezed bath}

As another application, consider a single mode subject to a squeezed bath. 
The Hamiltonian is  taken to be $H = \omega a^\dagger a$ and  the damping matrix is once again $\Gamma = \gamma I_2$.
The diffusion matrix $F$, on the other hand,  has the form $F = \Gamma Q_t$, with 
\begin{equation}\label{Q_squeezing}
    Q_t = \begin{pmatrix}
    N+\nicefrac{1}{2} & M e^{-2 i \omega_p t} \\[0.2cm]
    M^* e^{2i \omega_p t} & N+\nicefrac{1}{2}
    \end{pmatrix}.
\end{equation}
Here $N+\nicefrac{1}{2} = (\bar{n}+\nicefrac{1}{2}) \cosh 2r$ and $M = e^{i \theta} (\bar{n}+\nicefrac{1}{2}) \sinh 2r$ are related to the thermal occupation $\bar{n}$ and the squeezing parameter $z = r e^{i \theta}$.
The peculiar feature of the diffusion matrix~(\ref{Q_squeezing}) for the squeezed bath, is the explicit time-dependence, with frequency $\omega_p$.

Due to this time-dependence, the Lyapunov equation~(\ref{lyap}) will never reach a steady-state. 
Notwithstanding,  rotating frame CM $P \Theta P^\dagger$, where $P = \text{diag}(e^{i \omega_p t}, e^{-i \omega_p t})$, will obey a time-independent Lyapunov equation and will thus reach a unique steady-state. 
In the long-time limit, we get
\begin{equation}
    \Theta = \begin{pmatrix} 
    N+\nicefrac{1}{2} & \tilde{M}e^{-2 i \omega_p t} \\[0.2cm]
    \tilde{M}^* e^{2i \omega_p t} & N +\nicefrac{1}{2}
    \end{pmatrix},
\end{equation}
which is close to the bath-imposed CM~$Q_t$ in Eq.~(\ref{Q_squeezing}), with one fundamental difference: the squeezing parameter $M$ is modified to 
\begin{equation}
    \tilde{M} = \frac{\gamma }{\gamma + 2 i \Delta} M,
\end{equation}
where $\Delta = \omega_p - \omega$ is the detuning between the system frequency $\omega$ and the bath-imposed frequency $\omega_p$. 
The squeezing is therefore altered (reduced in magnitude and rotated) due to the presence of the detuning.

The flow matrix~(\ref{upsilon}) in this case becomes
\begin{equation}
    \Upsilon = \frac{\gamma |M|^2}{\nn^2}\begin{pmatrix}
    1-\eta & e^{-2 i \omega_p t} \frac{(N+\nicefrac{1}{2})}{M^*} (\eta-1) \\[0.2cm]
     e^{2 i \omega_p t} \frac{(N+\nicefrac{1}{2})}{M} (\eta^*-1) & 1- \eta^*
    \end{pmatrix},
\end{equation}
where $\eta = \gamma/(\gamma + 2 i \Delta)$. 
The flow matrix will therefore be non-zero provided there is a finite squeezing in the bath, $M\neq 0$ and there is a finite detuning $\eta \neq 1$. 
Taking the trace of this expression yields the entropy flux~(\ref{Pi_Phi_traces}), 
\begin{equation}\label{Phi_squeezed_bath}
    \Phi = \frac{4\gamma \Delta^2}{\gamma^2 + 4 \Delta^2} \sinh^2(2r).
\end{equation}
Even though $\Theta_t$ remains time-dependent, even in the long-time limit, the entropy~(\ref{entropy}) becomes time-independent so $dS/dt = 0$. As a consequence, Eq.~(\ref{Phi_squeezed_bath}) also represents the entropy production rate in the steady-state. 
We therefore see that, due to the detuning, the system is continuously producing some entropy $\Pi$, all of which flows to the bath ($\Phi$). 

\subsection{Two-mode squeezing interaction and local baths}

Finally, we consider two bosonic modes interacting with the Hamiltonian 
\begin{equation}
    \mathcal{H} = - \frac{i\chi}{2} (a^\dagger b^\dagger - a b). 
\end{equation}
We assume that both modes are connected to local environments at the same occupation $\bar{n}$, but with different damping rates $\gamma_a$ and $\gamma_b$. 
The Lyapunov equation~(\ref{lyap}) in this case is only stable provided $\chi^2 < \gamma_a \gamma_b$ (notice how this implies that a stable solution can only be reached when both $\gamma_{a,b} \neq 0$).
The flow matrix~(\ref{upsilon}) in the steady-state can written as 
\[
    \Upsilon = \frac{2\chi \gamma_a \gamma_b}{( \gamma_a+\gamma_b)(\gamma_a \gamma_b - \chi^2)}
    \begin{pmatrix}
    \chi & 0 & 0 & -\sqrt{\gamma_a \gamma_b} \\[0.2cm]
    0 & \chi & -\sqrt{\gamma_a \gamma_b} & 0 \\[0.2cm]
    0 & -\sqrt{\gamma_a \gamma_b} & \chi & 0 \\[0.2cm]
    -\sqrt{\gamma_a \gamma_b} & 0 & 0 & \chi.
    \end{pmatrix}.
\]
The flow matrix is seen to be independent of the bath temperature $\bar{n}$.
The corresponding entropy flux/production rate in the steady-state will then be 
\begin{equation}
    \Pi = \Phi = \frac{4 \gamma_a \gamma_b \chi^2}{(\gamma_a + \gamma_b)(\gamma_a \gamma_b - \chi^2)},
\end{equation}
which relies exclusively on the two-mode squeezing interaction $\chi$, being zero only if $\chi = 0$ (note that we cannot set $\gamma_{a,b} = 0$, without also setting $\chi = 0$, since this would lead to a unstable dynamics). 

\section{Discussions and conclusions}

Onsager's relation represent one of the most relevant results in the framework of non-equilibrium thermodynamics. 
However, they are usually restricted to linear response and thus are valid only close to equilibrium. 
No such relation holds in general for systems far from equilibrium. 
Similarly, there is also no general relation extending Onsager's relation to the quantum regime. 
In this paper we have shown that for Gaussian bosonic maps, exceptionally, it is possible to derive an Onsager relation which is valid arbitrarily far from equilibrium. 
This relation takes a matrix form and has a more complicated non-linear structure which reduces to the usual quadratic form of Onsager's original formula in the limit of linear response. 

Such a non-linear dependence emphasizes an asymmetry of  out-of-equilibrium processes, concerning the flow of entropy from the system to the bath and vice-versa. 
In the usual Onsager formulation, since the entropy production is quadratic, it depends only on the magnitude of the flow and not on its direction. 
Our results, however, show how far from equilibrium this is no longer true. 
Albeit restricted to the specific context of Gaussian states and Gaussian preserving maps, this provides an example of fundamental new features which may emerge as systems are driven far from equilibrium. 

\emph{Acknowledgements} GTL acknowledges the support from the S\~ao Paulo research foundation FAPESP under grants 2018/12813-0, 2017/50304-7 and 2017/07973-5.

\bibliography{library}
\end{document}